\documentclass[journal,twoside,web]{ieeecolor}
\usepackage{tmi}
\usepackage{cite}
\usepackage{amsmath,amssymb,amsfonts}
\usepackage{algorithmic}
\usepackage{graphicx}
\usepackage{textcomp}
\def\BibTeX{{\rm B\kern-.05em{\sc i\kern-.025em b}\kern-.08em
    T\kern-.1667em\lower.7ex\hbox{E}\kern-.125emX}}
\begin{document}
\title{Quasi-Real Time Multi-Frequency 3D Shear Wave Absolute Vibro-Elastography (S-WAVE) System for Prostate}
\author{Tajwar Abrar Aleef, Julio Lobo,~\IEEEmembership{Member,~IEEE,}
        Ali Baghani,~\IEEEmembership{Member,~IEEE,}
				Hani Eskandari,~\IEEEmembership{Member,~IEEE,} Hamid Moradi,
				Robert Rohling,~\IEEEmembership{Member,~IEEE,}
				S. Larry Goldenberg,
				William James Morris, S. Sara Mahdavi,
				and Septimiu E. Salcudean,~\IEEEmembership{Fellow,~IEEE}	
\thanks{This work was supported by the Canadian Institutes of Health Research (CIHR), Grant number CIHR MOP-1422439 and the C.A. Laszlo Chair held by Prof. Salcudean.}
\thanks{T. A. Aleef is with the School of Biomedical Engineering, University of British Columbia, Vancouver, BC, Canada, email: tajwaraleef@ece.ubc.ca}
\thanks{H. Moradi, R. Rohling, S. E. Salcudean are with the Department of Electrical and Computer Engineering, University of British Columbia, Vancouver, BC, Canada, email: tims@ece.ubc.ca}
\thanks{J. Lobo was with the Department of Electrical and Computer Engineering, University of British Columbia, Vancouver, BC, Canada. He is now with Sonic Incytes Medical Corp., Vancouver, BC, Canada}
\thanks{A. Baghani was with the Department of Electrical and Computer Engineering, University of British Columbia, Vancouver, BC, Canada. He is now with Salesforce, Toronto, ON, Canada}
\thanks{H. Eskandari was with the Department of Electrical and Computer Engineering, University of British Columbia, Vancouver, BC, Canada. He is now with Ladybug Robotics, Vancouver, BC, Canada}
\thanks{S. S. Mahdavi is with BC Cancer - Vancouver Centre, BC, Canada}
\thanks{S. L. Goldenberg is with the Department of Urological Sciences, University of British Columbia, Vancouver, BC, Canada}
\thanks{W. J. Morris was a Radiation Oncologist at BC Cancer - Vancouver Centre, BC, Canada. He has now retired.}
}

\maketitle

%242 words

\begin{abstract}
This article describes a novel quasi-real time system for quantitative and volumetric measurement of tissue elasticity in the prostate. Tissue elasticity is computed by using a local frequency estimator to measure the three dimensional local wavelengths of a steady-state shear wave within the prostate gland. The shear wave is created using a mechanical voice coil shaker which transmits multi-frequency vibrations transperineally. Radio frequency data is streamed directly from a BK Medical 8848 trans-rectal ultrasound transducer to an external computer where tissue displacement due to the excitation is measured using a speckle tracking algorithm. Bandpass sampling is used that eliminates the need for an ultra fast frame rate to track the tissue motion and allows for accurate reconstruction at a sampling frequency that is below the Nyquist rate. A roll motor with computer control is used to rotate the  sagittal array of the transducer and obtain the 3D data. Two CIRS phantoms were used to validate both the accuracy of the elasticity measurement as well as the functional feasibility of using the system for {\em in vivo} prostate imaging. The system has been used in two separate clinical studies as a method for cancer identification. The results, presented here, show numerical and visual correlations between our stiffness measurements and cancer likelihood as determined from pathology results. Initial published results using this system include an area under the receiver operating characteristic curve of 0.82$\pm$0.01 with regards to prostate cancer identification in the peripheral zone.
\end{abstract}
\begin{IEEEkeywords}
Ultrasound, prostate cancer, absolute vibro-elastography, shear wave elastography
\end{IEEEkeywords}

\section{Introduction}
\label{sec:introduction}
\IEEEPARstart{I}{n} the field of medical imaging, elastography has been introduced as a modality that can be used to measure and display the mechanical properties of tissue~\cite{lerner1988sono}. The additional information provided by elastography to conventional imaging can be used to help diagnose and guide treatment of a number of different medical conditions that include hypertension and fibrosis in organs~\cite{witters2009nliverfib,castera2005liver,warner2011kidney}, fetal development in placenta~\cite{edwards2020use}, as well as various cancers~\cite{correas2015prostate,arroyo2018breast, sang2017accuracy,eskandari2008viscoelastic}.

Several different methods have been proposed in the field of ultrasound elastography~\cite{sigrist2017ultrasound} and the trend has been to migrate from relative, quasi-static techniques~\cite{ophir1991elastography,lorenz1999new} to absolute, quantitative techniques~\cite{wildeboer2020automated,dai2021correlation,ji2019stiffness,zhang2008quantitative,zhai2010acoustic,ahmad2013transrectal,gennisson2015} which reduce user dependence and increase repeatability~\cite{ahmad2013transrectal}. Most of these systems accurately measure the transient speed of a shear wave generated using an acoustic radiation force-- where this force is generated by the transducer sending a high power focused beam to the tissue. The speed of this wave is proportional to the tissue elasticity. The main drawback of these methods however is that there is a limit imposed on the amplitude of the acoustic impulse which restricts the maximum signal to noise ratio that can be achieved. This limitation arises from the fact that acoustic impulses can lead to tissue heating which can harm the patient~\cite{nightingale2011acoustic}. As a result, strict FDA guidelines control the amplitude and duration of pulses that can be used. The issue is especially restrictive if continuous or repeated imaging is required, for instance, as needed in real-time guided interventions. Deep tissue excitation and full volume acquisitions are also not always possible with such an approach.

To overcome these limitations, our group has previously developed a new kind of quantitative elasticity measurement technique called Shear Wave Absolute Vibro-Elastography (S-WAVE).
The technique uses multi-frequency, steady-state external excitation, and measurement of the shear wave displacement field over a volume \cite{baghani2012real,schneider2012remote}.
The elasticity modulus is computed either by solving an inverse FEM~\cite{honarvar2014vibro} problem or by measuring the wavelength of a three-dimensional steady-state shear wave field~\cite{manduca1996local}. 
The benefit of multi-frequency elastography has been demonstrated in MR elastography~\cite{asbach2008assessment} for non-real time applications as well as in shear wave speed methods~\cite{deffieux2009shear}. 
Different hardware and different imaging settings have been used for specific organ imaging, including {\em in vivo} liver~\cite{zeng2020three} imaging, {\em ex vivo} placenta~\cite{abeysekera2017swave} imaging, and breast elastography~\cite{shao2021breast}.
In this article we propose and characterize a prostate S-WAVE system that can perform quasi-real time \& 3D volumetric imaging and can produce images at the depth of conventional ultrasound without causing tissue heating. Initial findings with this setup have previously been summarized in Lobo et al.~\cite{lobo2015prostate} and are expanded here with further system characterization and details. 
The system described here is the first quasi-real time prostate 3D quantitative elastography system, with the following contributions:\\
1) The use of a bandpass sampling technique~\cite{eskandari2011bandpass} to allow tracking of high frequency tissue motion with an ultrasound frame rate that is well below the Nyquist rate. This removes the need for either a programmable transducer crystal firing sequence to increase the frame rate~\cite{baghani2010high} or an expensive parallel receive ultrasound system.
This also enables Radio frequency (RF) data from the ultrasound machine to be streamed to an external PC through a frame grabber card, with the machine operating in conventional B-mode.\\
2) Motorized volumetric acquisition with a commonly used 2D transducer without changing the volume boundary conditions; the only 3D transducer currently available for prostate imaging is an end-firing  wobbler that substantially deforms the prostate making it difficult to correlate histology results with imaging.\\
%or a 2D matrix array~\cite{gennisson2015}. 
3) Non-invasive transperineal excitation to induce shear waves to the prostate that have been reported to produce repeatable and consistent results with magnetic resonance elastography~\cite{sahebjavaher2013transperineal}.\\ 
4) Validation in tissue-mimicking phantoms, over a range of frequencies.\\
5) A comparison of elasticity measurements with whole-gland sampling biopsies acquired during a focal brachytherapy clinical study. This complements an earlier study that compared histology results from excised prostates following prostatectomy procedures with S-WAVE images and reported in conference proceedings~\cite{mohareri2014multi}. The results from these studies will be summarized in Section \ref{results}.\\
 The contributions of this article can be used in any generic system developed for quasi-real time 3D absolute elastography, which can be stand alone and independent of the ultrasound machine or probe.
 
The following sections of the article are organized as follows: Section \ref{methods} covers the hardware setup for the overall system and the signal processing pipeline for S-WAVE. Section \ref{validation setup} and \ref{results} covers the details and results from phantom validation and {\em in vivo} patient data from the two different clinical studies. Lastly, Section \ref{conclusions} discusses the results, limitations, future improvements and concludes the article.

\section{Methods}

\label{methods}

\subsection{Hardware Setup For Prostate Imaging}
\label{sec:ProstateSystem}

\begin{figure}[!t]
\centering
      \includegraphics[width=3.3in, trim={0.3cm 2cm 3.6cm 0},clip]{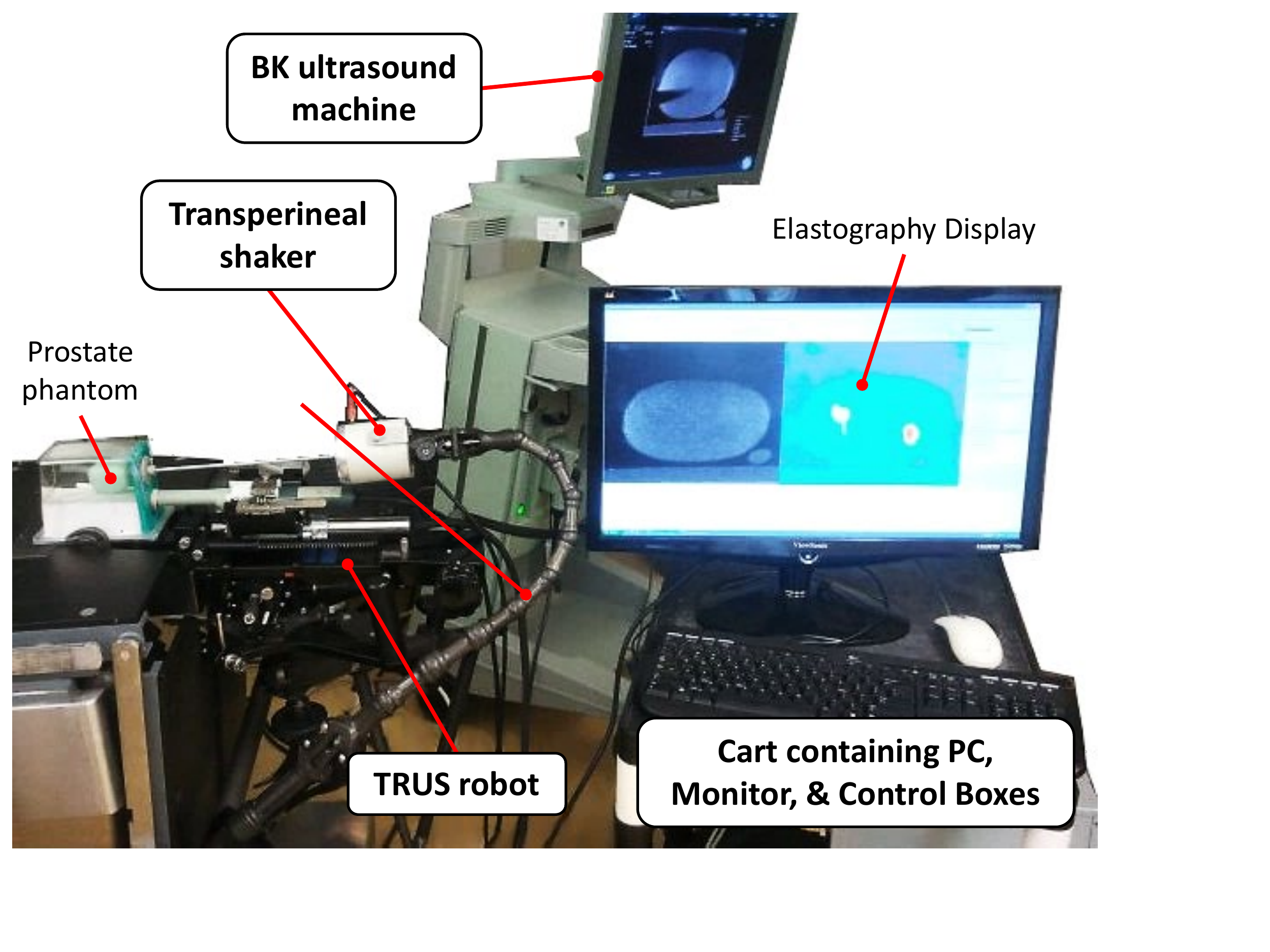}\\ %{<left> <lower> <right> <upper>}
 \caption{The full prostate S-WAVE system. The four main hardware components are labelled in bold. The cart contains the PC with monitor and the two control boxes for the roll motor and the transperineal shaker.}
 \label{fig:full_setup}
\end{figure}

\begin{figure}[!t]
\centering
      \includegraphics[width=3.5in, trim={0 11cm 12cm 0},clip]{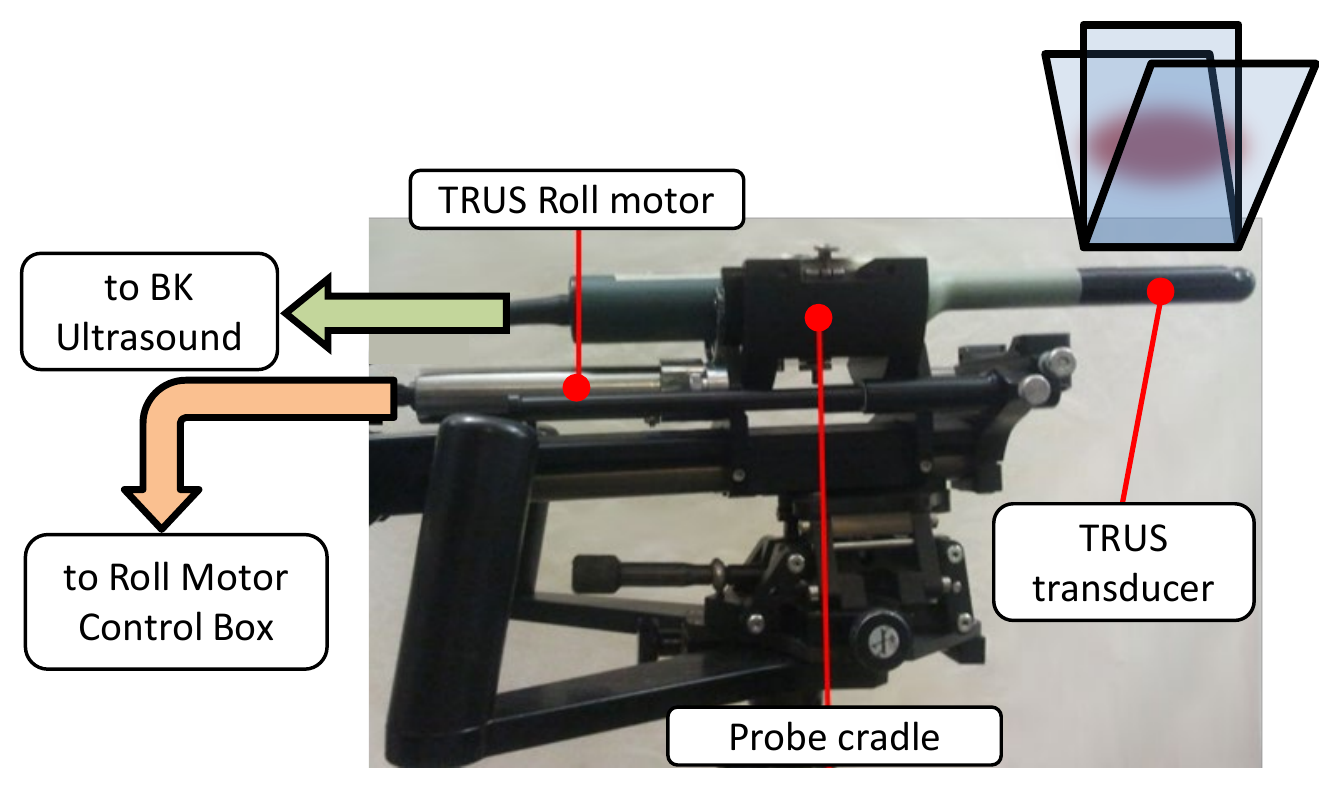}
 \caption{The TRUS robot with the BK transducer. An illustration of some of the sagittal imaging planes is also shown. } 
 \label{fig:trus}
\end{figure}

This section describes the specific implementation of S-WAVE hardware for prostate imaging. The main hardware, seen in Fig.~\ref{fig:full_setup}, consists of the ultrasound machine with a transrectal ultrasound (TRUS) transducer to acquire raw RF data, a TRUS robot to roll the transducer and obtain a 3D volume, a transperineal shaker to provide mechanical excitation, and an external PC for computing elastography and controlling the shaker \& the TRUS robot. Each of the system components is described next. 

\subsubsection{Ultrasound Machine}
A BK Pro Focus ultrasound machine (BK Medical, Herlev, Denmark) is used with the BK 8848 4-12 MHz TRUS transducer. From the ultrasound, echo data in the form of in-phase/quadrature (IQ) modulated data is streamed to an external PC with a DALSA Xcelera-CL PX4 Full frame grabber card (Teledyne DALSA, Waterloo, Canada). On the external PC side, RF lines are reconstructed from the IQ data by reversing the modulation process. 

\subsubsection{TRUS Robot}
A separate control system called the TRUS robot is used to sweep the 2D transducer in order to obtain 3D RF data. The BK 8848 transducer probe is attached to a conventional brachytherapy TRUS stand and stepper (Micro-Touch$^{TM}$ and EXII Stepper (CIVCO Medical Solutions, Kalona, USA)). This stepper uses an encoder to read the ``roll'' angle of the probe, which is useful for needle insertion monitoring during transperineal biopsy and brachytherapy intervention.  
We designed a specific computer-controlled ``roll'' motor, with encoder, that replaces the conventional manually-rotated stepper, without substantially modifying the hardware and allowing our software to accurately control the angle of the probe about its long axis. Fig.~\ref{fig:trus} shows a side view of the TRUS robot with the motorized TRUS transducer held in the cradle.

\begin{figure*}[!t]
\centering
      \includegraphics[width=\textwidth, trim={0.3cm 6cm 4.2cm 0},clip]{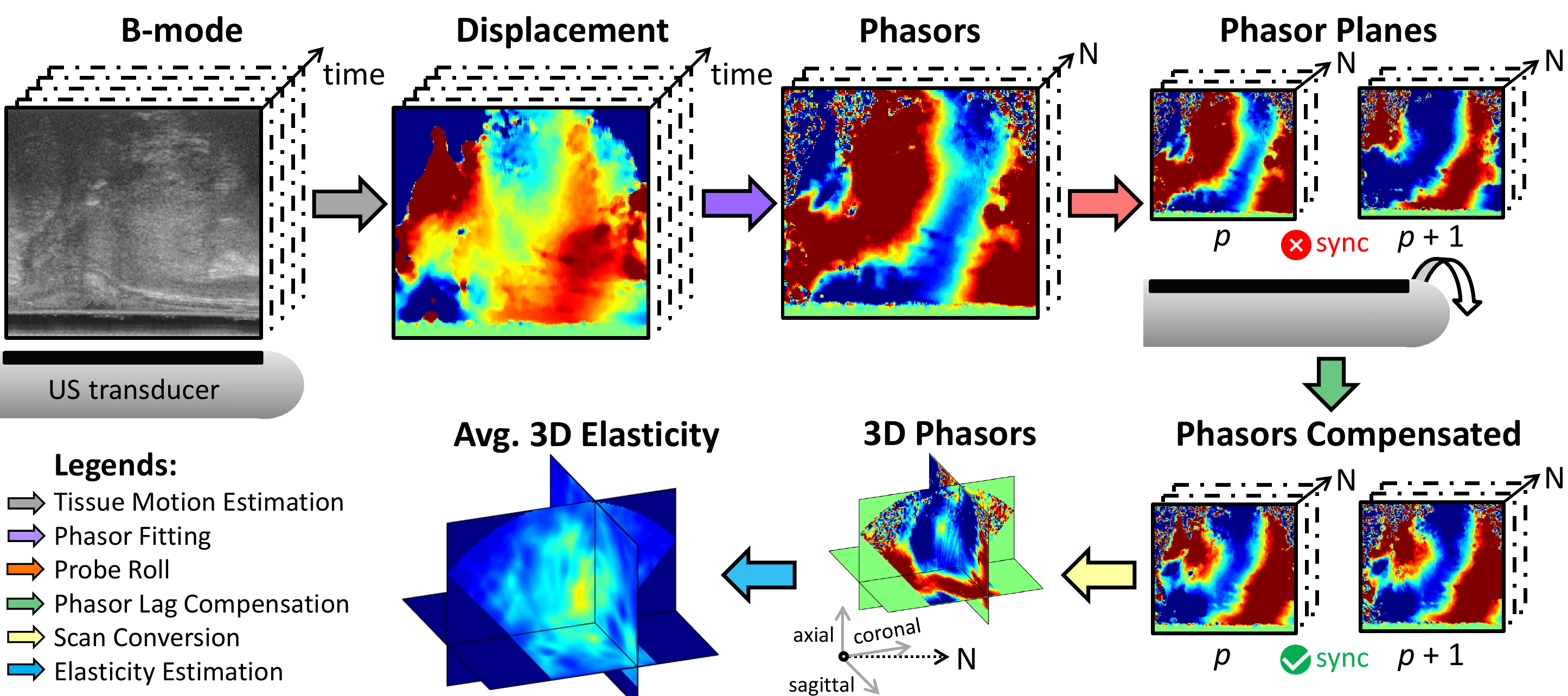}
 \caption{S-WAVE processing pipeline showing the steps involved in calculating the absolute elasticity for patient 1 from Focal Therapy Study. Raw RF time-series data is used to estimate tissue displacement generated from the shear waves. Phasors are found for each of the excitation frequencies (total $N$ frequencies) from the displacement data. 3D data is collected by rolling the transducer. Phasor planes are not in sync due to the presence of time delay from one plane ($p$) acquisition to the next ($p+1$). This lag is then compensated and the phasor planes are scan converted to generate 3D phasors for each excitation frequency. Lastly, the average absolute elasticity volumes are generated from the 3D phasors.}
 \label{fig:fullflow}
\end{figure*} 

\subsubsection{Transperineal Shaker}
Multi-frequency excitation is applied transperineally using a voice coil actuator. A circular shaped rod is connected to this shaker that makes contact with the perineum of the patient and generates forces/displacements in the inferior-anterior direction. The shaker can be mounted onto the patient's table by using a CIVCO Assist$^{TM}$ flexible arm. The flexible arm is maneuvered in order to provide good contact with the perineum; once that is achieved, the arm is locked into position using its handle bars. Fig.~\ref{fig:full_setup} displays the shaker together with the CIVCO arm and the circular rod. An Agilent U2761A function generator (Agilent Technologies, Santa Clara, USA), controlled by the external PC, is used to output the desired excitation frequencies. The excitation signal, $s(t)$, is the sum of multiple sinusoids:

\begin{equation}
\label{eq1}
s(t) =  \sum_{n=1}^{N} a_n\sin{(\omega_n t)}
\end{equation}

where $N$ is the number of frequency components each having amplitude $a_n$ and frequency $\omega_n$. The higher the $\omega_n$, the higher the attenuation of the signal with depth. To compensate for this, higher $a_n$ is set for higher $\omega_n$.

\subsection{Signal Processing Pipeline for S-WAVE} 
\subsubsection{Bandpass Sampling}
There are two viable methods of sampling tissue motion at high frequencies-- one that uses Doppler/Angiography-like sequences to acquire tissue motion in small sectors~\cite{baghani2010high} and one that uses bandpass sampling~\cite{eskandari2011bandpass}. The bandpass sampling method does not require users to program the ultrasound machine sequencer and therefore works with any generic ultrasound machine that can dump RF lines to an external link-- even when the machine is in clinical mode. The native frame rate of focused B-mode acquisition, due to speed-of-sound limitations, is of the order of 20-60~Hz for typically sized images. At this low frame rate, it is not possible to sample tissue with higher frequency tissue motion, at 70 Hz and above, as required to produce small enough shear wavelengths to fit within the region of interest and give good resolution of prostate elastography images. Bandpass sampling allows reconstruction of phasors of motion with frequency higher than the Nyquist frequency. This is possible as the excitation frequency and the sampling frequency ($f_s$) are known. Due to having a low $f_s$, the reconstructed phasor will be aliased and appear with a lower frequency. If $f_s$ satisfies (\ref{bandpass}), the reconstructed signal will resemble the original spectrum in the measurable baseband.

\begin{equation}
    \label{bandpass}
    \frac{2f_c+B}{m+1} \leq f_s \leq \frac{2f_c-B}{m}
\end{equation}

where, $f_c$ is the center frequency of the signal, $B$ is the bandwidth, and $m$ is the integer number of spectral half-shifts ($f_s/2$) needed to map the original spectrum in the baseband. For multiple excitation frequencies, $B$ of 10~Hz around every frequency component is a good choice to avoid overlap in the baseband. 

\subsubsection{Image Acquisition}
The volume acquisition consists of a programmed scan of the TRUS covering a rotation from -49$^\circ$ to 50$^\circ$ with 1$^\circ$ increments, acquiring 100 discrete imaging planes each with $M=20$ sagittal frames of RF data. The full 100-plane volume sweep takes $\sim$90 seconds to complete. The imaging depth and width for each plane are set to 55.2~mm and 60~mm, respectively, which are the preset values for prostate imaging. Each RF frame contains 214 scan lines that are sampled at 50~MHz. The RF frames are captured using the maximum achievable sampling rate at this depth and width, which is $f_s=42.66$ frames per second for this ultrasound machine. Following Bandpass sampling method, shear wave excitation frequencies are therefore selected so that they are not multiples of $f_s=42.66$~Hz or $f_s/2=21.33$~Hz and such that they do not overlap in the baseband (0 - $f_s/2=21.33$~Hz).

 \subsubsection{Tissue Motion Estimation}
A speckle tracking algorithm~\cite{zahiri2006motion} is used to estimate relative tissue displacement as a function of time. In this algorithm, scan lines are first divided into multiple overlapping windows with a window size of 1~mm and spatial overlap of 84\%. Next, 1D axial displacements are calculated by finding the maximum normalized cross-correlation (NCC) between these time-series windows using cosine-fitting for sub-sample displacement resolution. This generates a displacement time series image for each imaging plane.
%containing tissue motion generated by the steady-state shear wave. 

\subsubsection{Phasor Fitting}
Assuming a linear tissue response, at each voxel, the displacement $d(t_i)$ is the sum of complex phasors $U_{1:N}$ with excitation frequencies $\omega_{1:N}$ (Equation~\ref{eq2}). 

\begin{align}
     d(t_i) = \sum_{n=1}^{N} Re[~\overbrace{A_n e^{j\theta_n}}^{U_n} e^{j\omega_n t_i}~] \label{eq2}
\end{align}

where, $A_n$ and $\theta_n$ are the amplitude and phase of phasor $U_n$. $t_i$ is known from the ultrasound's scan sequence delay table and $\omega_n$ is the know excitation frequency. 

To recover the best fit for $U_n$, only its $A_n$ and $\theta_n$ needs to be calculated. By representing (\ref{eq2}) in its trigonometric form and taking the real part, the unknown and known parameters of $U_n$ can be expressed in matrix form of $x$ and $A$ of size $(N+1, 1)$ and $(M,  N+1)$, respectively (since there are $M=20$ frames, $i = 1:M$). 
Similarly, $d(t_i)$ is represented as a column vector $b$ of size $M$.\\
\begin{align}
     d(t_i) = & \sum_{n=1}^{N}[~\overbrace{A_n\cos\theta_n}^{x^1_n} \underbrace{\cos\omega_n t_i}_{a^1_n(t_i)} - \overbrace{A_n\sin\theta_n}^{x^2_n} \underbrace{\sin\omega_n t_i}_{a^2_n(t_i)}~] \label{eq3}\\
     \overbrace{d(t_i)}^{b_i} = & \overbrace{[\begin{smallmatrix} a^1_1(t_i) & -a^2_1(t_i) & a^1_2(t_i) & -a^2_2(t_i) & \cdots & a^1_N(t_i) & -a^2_N(t_i) & 1\end{smallmatrix} ]}^{A(i,:)} \nonumber \\ 
     & \times \underbrace{[\begin{smallmatrix} x^1_1 & x^2_1 & x^1_2 & x^2_2 & . & . & x^1_N & x^2_N & c\end{smallmatrix}]^T}_x 
\end{align}
where $c$ accounts for any DC offset. 
 
Finally, an estimate $\hat{x}$ of $x$ can be approximated for every voxel using a least-square fit to (\ref{eq5})
\begin{eqnarray}
\left[ 
\begin{array}{c}
b_1 \\
\vdots \\
b_M
\end{array}
\right]
=
\left[ 
\begin{array}{c}
A(1,:) \\
\vdots \\
A(M,:)
\end{array}
\right]
x
%\left[ 
%\begin{array}{c}
%x_1^1 \\
%x_1^2 \\
%\vdots\\
%x_N^1 \\
%x_N^2 \\
%1
%\end{array}
%\right]
\,\,\, \Leftrightarrow \,\,\, b = Ax
\label{eq5}
\end{eqnarray}
\begin{align}     
     \hat{x} = (A^TA)^{-1}A^Tb \label{pinverse}
\end{align}

From $\hat{x}$, the amplitude \& phase and therefore the phasor at each excitation frequency for every voxel is found which is then used to form 2D phasor images at each plane. These images contains a 2D projection of the steady-state shear wave as can be seen in Fig.~\ref{fig:fullflow}. 

\subsubsection{Phase Lag Compensation}
As each plane is collected at a different time (time delay as the probe rotates), the displacement data must be shifted in time to represent an equivalent simultaneous measurement. The time delays  ($\Delta T_p$) of each plane ($p$) with respect to the data capture start are recorded. Following \cite{baghani2010high}, the delay of each plane is compensated by multiplying the phasor $U_n(p)$ with a complex exponential: 
\begin{equation}
    U_n^{sync}(p) = U_n(p) e^{-j\omega_n \Delta T_p} \,\,\, 
\end{equation}
The set of \textit{synchronized} 2D displacement phasor images can then be scan-converted based on transducer specific parameters to get 3D phasor volumes for each excitation frequency. 

\subsubsection{Elasticity Estimation}
\label{sec:algorithm}
It can be shown that dynamic and external excitation at a given temporal frequency, $\omega_{n}$, produces steady-state ``compression'' and ``shear'' waves within a given material~\cite{kolsky1964stress}. 
%baghani2011travelling}. 
The spatial frequencies, or wavenumbers (angular), of the compression wave, $k_{c_n}$, and the shear wave, $k_{s_n}$, are functions of the Lam\'{e} parameters of the material which describe its viscoelastic properties. 
$ks_n$ can be approximated using lognormal quadrature filters as Local Frequency Estimators (LFE)~\cite{knutsson1994local,manduca1996local}. 
A $3^{rd}$ order Butterworth bandpass filter is applied to the 3D phasor volumes in the frequency domain prior to applying LFE. This reduces motion artefacts and noise in the phasors. The cutoff frequencies for this bandpass filter are set to remove frequency components that corresponds to elasticity measurement outside the $[1, 85]$ kPa.\\
The LFE algorithm involves using directional filters followed by a bank of frequency filters to obtain an estimate for $k_{s_n} $ at every voxel for each $\omega_n$ from the filtered phasor volumes; the corresponding elasticity at frequency $\omega_n$ is computed as $E_n = 3\rho ( \omega_n / k_{s_n} )^2$, with $\rho=1000 kg/m^3$ (as soft tissues density can be estimated to be similar to that of water).
Since the excitation frequencies are close to each other we ignore the frequency dependence and compute the average elasticity at a given voxel as:
\begin{equation}
    E = 3\rho {\frac{1}{N}\sum_{n=1}^{N}\big(\frac{\omega_n}{k_{s_n}}\big)^2} \label{eq7}
\end{equation}
 
Averaging out the elasticity estimated from different $\omega_n$ reduces artifacts that can arise during displacement estimation.

All of the above algorithms are implemented to run on a CUDA enabled GPU. This allows parallel estimation of tissue motion and therefore makes quasi-real time elasticity estimation possible. Each step of the signal processing pipeline is illustrated in Fig.~\ref{fig:fullflow}.

\section{Validation Setup}
\label{validation setup}
\subsection{Validation on Phantoms}

Two CIRS (Computerized Imaging Reference Systems, Norfolk, USA) elasticity phantoms were used to validate the accuracy of the elasticity measurement as well as the feasibility of using the full system in a clinical environment.

The measurement values were validated using the CIRS model 049 phantom which has inclusions inside a uniform background material. The manufacturer's datasheet provides elasticity values for each inclusion as well as the background. Separately, Baghani et al.~\cite{baghani2011travelling} have performed Magnetic Resonance Elasticity (MRE) measurements on the same phantom and have published the elasticity value of its background and lesions. The CIRS model 049 phantom does not have a simulated rectum and so the measurements were made by turning the TRUS robot system upside down and sweeping a volume from above the phantom. But the same imaging parameters were used for both phantom studies to make a fair comparison to the original system. The elasticity values measured for these inclusions (identified as ``hard inclusion'' and ``medium inclusion'' in the remainder of this manuscript) as well as the background were then compared to values provided by the manufacturer as well as to the MRE measurements.

To demonstrate the feasibility of using the system for prostate elasticity, where the TRUS system could be oriented similarly to that of a clinical patient setting, a CIRS model 066 prostate phantom was used. This phantom, seen in Fig.~\ref{fig:full_setup}, is a simulated prostate that contains stiffer inclusions quoted as being approximately 3 times stiffer than the simulated prostate. The rectum and perineum are both also simulated and so the entire system as described in Section~\ref{sec:ProstateSystem} can be tested. Although an exact value for this elastic modulus is not given, based on the manufacturer, it can be estimated to be close to real prostate tissue. Zhang et al.~\cite{zhang2008quantitative} have previously determined normal prostate tissue to have an elastic modulus of $15.9\pm5.9$~kPa so we estimate this value for the simulated prostate tissue and $47.7\pm17.7$~kPa (3 times stiffer) for the hard inclusions.

For the phantoms, a multi-frequency signal with 144~Hz, 165~Hz and 181~Hz components was used. With the sampling rate, $f_s=42.66$, the frequencies are measured as 16.0~Hz ($|144-3f_s|$), 5.4~Hz ($|165-4f_s|$) and 10.4~Hz ($|181-4f_s|$) in the baseband and are therefore far enough apart for accurate frequency reconstruction.

\subsection{Patient Data Collection}

This section describes the data collection procedure for two clinical studies that were used to validate the proposed system.

\subsubsection{Prostatectomy Study}
\label{sec:prostatectomyStudy}
 
In this institutionally approved study, signed consent was obtained from ten patients with clinically organ-confined prostate cancer undergoing robot-assisted radical prostatectomy at Vancouver General Hospital (Vancouver, Canada). As this was the first clinical trial with this system, no prior information of the ideal excitation frequency range for {\em in vivo} prostate was known. Ideally, high frequency excitation has the benefit of more robust $k_s$ estimation but with the drawback of poorer penetration depth. Therefore to find the ideal excitation frequency range, before the prostatectomy, we captured several S-WAVE volumes from each patient with frequencies which ranged from 75~Hz to 180~Hz. To reduce the time required to collect all the data in the operating room, we chose a frequency step of 2.5~Hz (rounded to the nearest integer as required by our function generator). Data was then collected at successive excitation frequencies until shear waves were no longer visible-- indicating the frequency limit after which shear waves are not reaching the prostate. 
Right after the prostatectomy, the excised prostate was sent to pathology, where whole-mount histopathology slides were obtained by evenly slicing the gland. A pathologist at the Vancouver General Hospital then examined slices of the excised prostate glands and outlined any cancerous regions, assigning a Gleason score for each region. In order to compare the elastography results to histopathology slices, which are transverse slices, the elastography volume must first be interpolated into a cartesian grid from the stack of sagittal slices. A slice-to-surface, particle-filter-based registration technique~\cite{nir2013registration} is then used to match transverse planes in the elastography volume to corresponding histopathology slices. These registered whole-mount histopathology images can then be used as the ground truth for evaluating cancer detection capability of our prostate S-WAVE system. A specific data analysis approach to the the multiparametric nature of the images collected with this system has been presented in~\cite{mohareri2014multi}. In that work, the absolute, real and imaginary values of the phasor images as well as the quantitative elasticity values were all used to train a random forest classifier to identify cancer.

\begin{table*}[!t]
%\renewcommand{\arraystretch}{1.3}
% if using array.sty, it might be a good idea to tweak the value of
% \extrarowheight as needed to properly center the text within the cells
\caption{Comparison of elasticity values for different regions and measurement techniques on CIRS 049 phantom} %COMPARISON OF ELASTICITY MEASRUEMENTS FOR DIFFERENT REGIONS AND MEASUREMENT TECHNIQUES
\label{tab:ElastResults}
\centering
% Some packages, such as MDW tools, offer better commands for making tables
% than the plain LaTeX2e tabular which is used here.
%\[
\begin{tabular}{  c | c | c | c  }
%\hline
	\bf{Measurement Technique}                                                  & \bf{Hard Inclusion (kPa)}   & \bf{Medium Inclusion (kPa)} & \bf{Background (kPa)} \\ \hline
	{S-WAVE (mean$\pm$std)}                     & 50.0$\pm$7.9       & 37.5$\pm$8.2       & 22.2$\pm$5.5 \\ \hline
	{Magnetic Resonance Elasticity~\cite{baghani2011travelling} (mean$\pm$std)} & 49.4$\pm$16.9      & 36.4$\pm$2.0       & 22.3$\pm$3.0 \\ \hline
	{Manufacturer's Datasheet Values}                                           & 62                 & 54                 & 29         \\ 
\end{tabular}
%\]
\end{table*}

\subsubsection{Focal Therapy Study}
\label{sec:biopsyStudy}

To investigate the feasibility of focal therapy in low-dose-rate prostate brachytherapy (LDR-PB) for early stage prostate cancer, patients were recruited in an institutionally approved pilot study %(ClinicalTrials.gov Identifier: NCT01830166) 
at the BC Cancer (Vancouver, Canada). 
Patients underwent whole-gland template mapping biopsies to determine the size and location of any tumours within the gland. Based on the size of the prostate, following a needle template (see Fig.~\ref{fig:FTCPpat5}), 20-50 transperineal biopsy samples were obtained with the patient under general anaesthetic. Just prior to the biopsy procedure, after the patient was asleep, a S-WAVE volume was collected. Observations from the Prostatectomy study (Section \ref{sec:prostatectomyStudy}) showed that consistent wave penetration was possible for all frequencies lower than 90~Hz. With the frequency range now known, as with the phantom study, a multi-frequency signal was selected such that components were evenly spaced in the baseband which allows for accurate bandpass reconstruction.
A multi-frequency signal with a lower component range of 69~Hz, 75~Hz and 80~Hz (which correspond to baseband frequencies of 16.3~Hz ($|69-2f_s|$), 10.3~Hz ($|75-2f_s|$) and 5.3~Hz ($|80-2f_s|$), respectively) was used to provide adequate shear wave penetration. During the biopsy procedure, each core was deposited into separate labelled containers and sent to pathology for hemotoxylin and eosin (H\&E) staining and cancer identification. 
An expert pathologist from BC Cancer then reported the presence and location of cancer within the cores. Using an automatic transperineal biopsy registration technique~\cite{aleef2022}, following the needle template, pathology of the cores were mapped to the B-mode and S-WAVE volumes. An extended cancer correlation analysis has not yet been performed, the complete study will involve using classification methods similar to those used in the Prostatectomy study (Section \ref{sec:prostatectomyStudy}) to compute the accuracy of using this system to detect cancerous tissue by comparing them to the biopsy results.

\begin{figure}[!t]
\centering
      \includegraphics[width=3.3in] {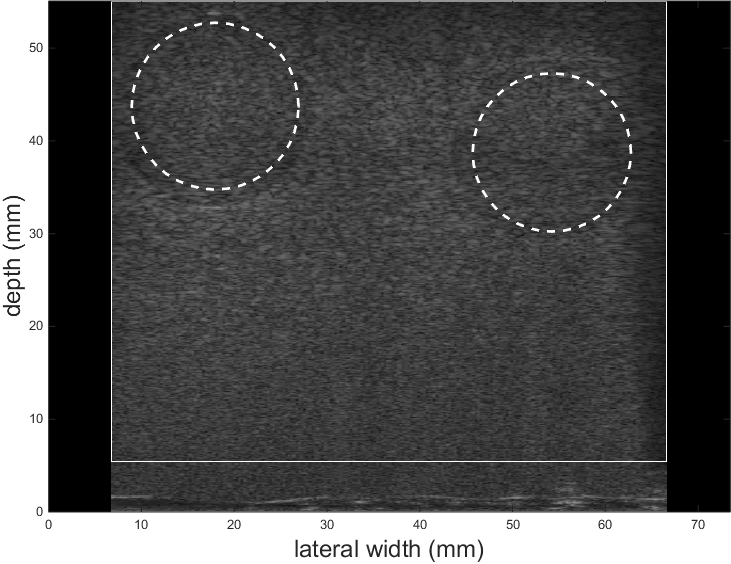}\\
 \caption{B-mode image showing the regions used to average the elasticity for the two inclusions (dotted lines) from CIRS 049 phantom. The solid white box is the region used for the background.}
 \label{fig:cirs049ROI}
\end{figure}

\begin{figure}[!t]
\centering
      \includegraphics[width=3.3in] {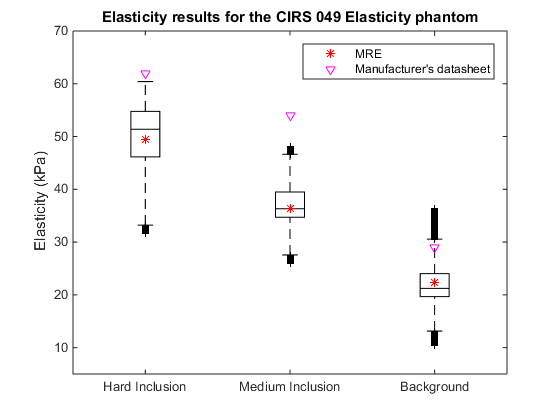}\\
 \caption{Elasticity value of the inclusions and background of the CIRS 049 phantom obtained from our prostate S-WAVE system, MRE, and Manufacture provided datasheet. The box plot represents the median, interquartile range, max/min, and outliers of elasticity measured using our S-WAVE system for the selected regions.}
 \label{fig:cirs049results}
\end{figure}

\begin{figure*}[!t]
\centering
      \includegraphics[width=\textwidth, trim={0 6.5cm 2.5cm 0}, clip] {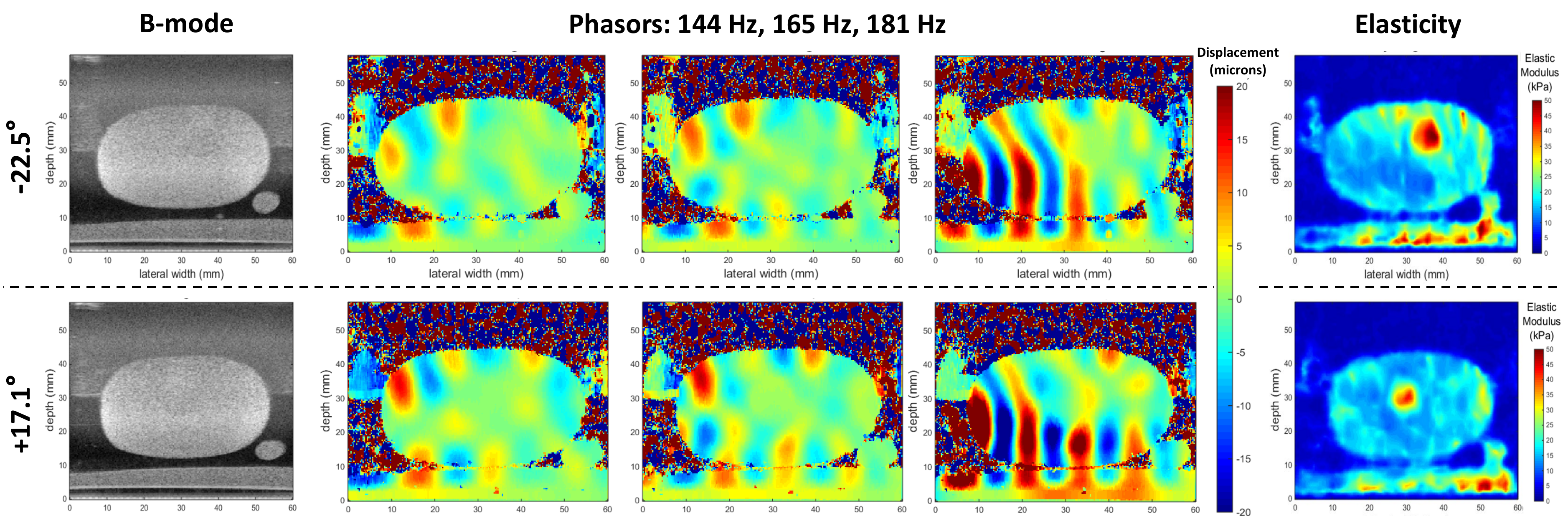}\\
 \caption{For roll angle of -22.5$^\circ$ and +17.1$^\circ$, this figure shows the B-mode, Multi-frequency phasors (144~Hz, 165~Hz, and 181~Hz), and Absolute Elasticity results from CIRS 066 prostate phantom. The red salt-and-pepper noise in the phasors result from the fact that there are no scatterers present in the region surrounding the simulated prostate. The scatterers produce the speckles in an ultrasound image that are needed for the tissue motion tracking. Positive identification of two hard inclusions can be seen from the two absolute elasticity images.}
 \label{fig:phasDisp}
\end{figure*}

\section{Results}
\label{results}

As a demonstration of the full system both phasor displacement and absolute elasticity images are presented from the CIRS 066 prostate phantom. For the patient studies, interpolated transverse absolute elasticity images are displayed side-by-side with their corresponding pathology. 
  
\subsection{Phantom Results}

In order to validate the Young’s modulus measurement, the system was used to measure the elasticity of two inclusions and the background material of the CIRS 049 phantom. The measured values were then compared to the manufacturer’s datasheet values as well as MRE measurements of the same phantom. Fig.~\ref{fig:cirs049ROI} shows a B-mode plane of this phantom with the regions that was selected to average the elasticity values for the two inclusions and background. The results are summarized in Table~\ref{tab:ElastResults} and illustrated graphically in Fig.~\ref{fig:cirs049results}.

The results show that the elasticity values match very closely with MRE measurements. As described in the MRE study, the differences between the quantitative prostate vibro-elasticity results and the manufacturer's datasheet values are likely due to issues with aging and hydration of the phantom and the controlled temperature at which the measurements were made~\cite{baghani2011travelling}. Nevertheless, the values for all three measurement techniques show the same trend.

\begin{figure*}[!t]
\centering
      \includegraphics[width=\textwidth, trim={0 9.5cm 7.5cm 0}, clip] {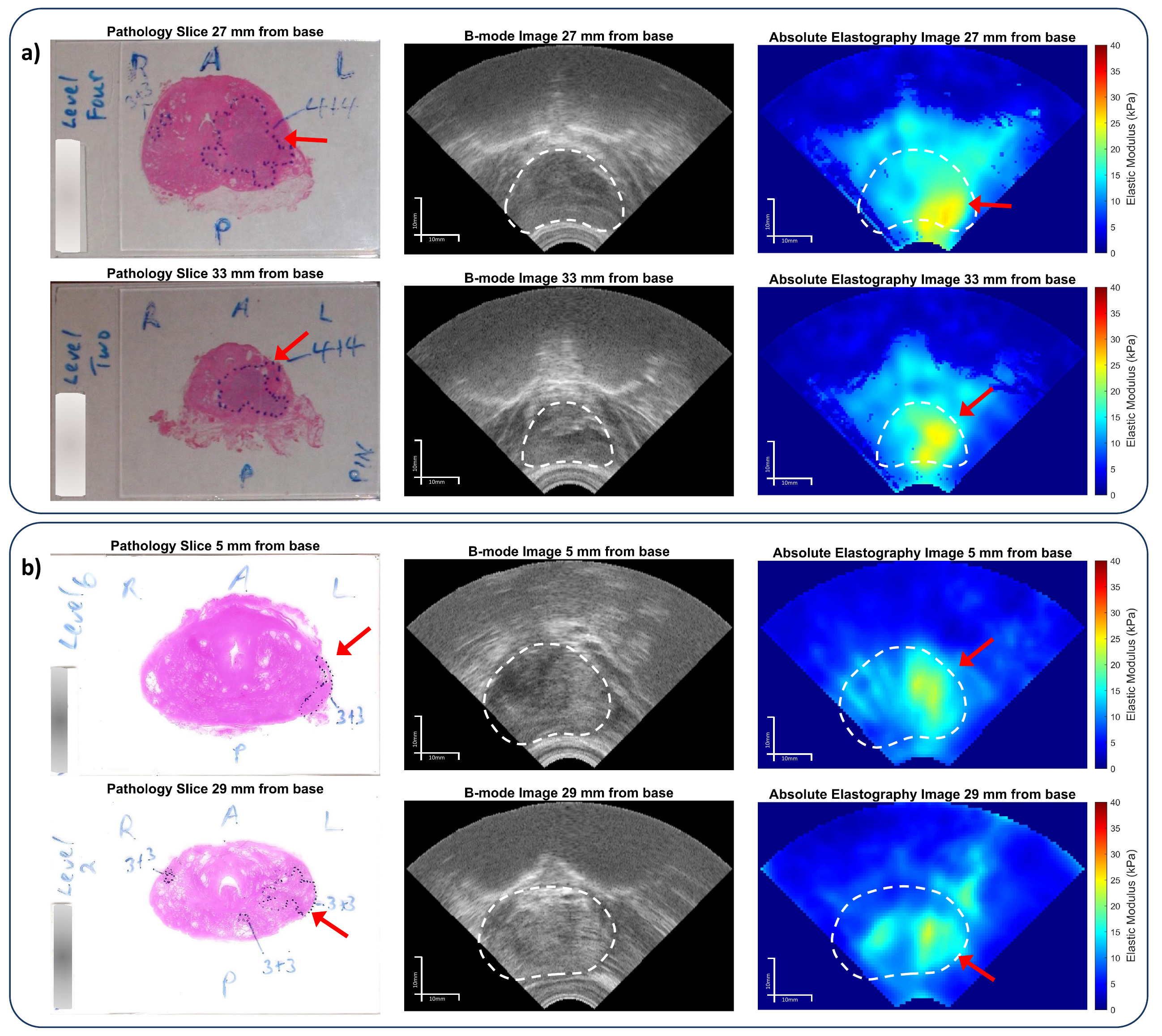}\\
 \caption{The comparison of histopathology images showing positively identified cancerous regions within the dotted blue lines (left) to B-mode images (middle) and the absolute elasticity values (right) for different slices in patient 6 (a) and patient 7 (b) from the Prostatectomy Study.}
 \label{fig:prostResults6}
\end{figure*}

The CIRS 066 prostate elasticity phantom was used to replicate a clinical setup and verify the feasibility of using the entire system as intended to measure the Young’s modulus of the prostate gland. This was done by computing absolute elasticity maps of the phantom for the entire volume. In all the images, the horizontal axis runs from inferior to superior (i.e. the apex of the prostate phantom is located on the left side of the sagittal image and base on the right). The vertical axis is the axial line of the ultrasound image and runs from the probe (at the bottom of the image) upward.

To demonstrate the intermediary step of detecting the shear wave within the prostate phantom, the phasor displacement image (in microns) at plane 26 and 70 (which corresponds to a roll angle of -22.5$^\circ$ and +17.1$^\circ$) for each frequency is shown in Fig.~\ref{fig:phasDisp}. Each image shows the projected steady-state shear wave. The local spatial wavelengths of this wave can be seen to get smaller for higher temporal excitation frequencies. Also, since inclusions are also visible in these planes (compare with elasticity images in Fig.~\ref{fig:phasDisp}), the reader may be able to see the increase in wavelength seen in the stiffer regions.

Fig.~\ref{fig:phasDisp} also shows the absolute elastography results in kPa for planes for these two roll angles. The elastic modulus was computed by averaging the results from transperineal excitation at 144~Hz, 165~Hz and 181~Hz. It can be seen from the figure that the estimated elasticity calculated for the hard inclusions and the surrounding simulated tissue are around $42$~kPa and $18$~kPa, respectively. This falls within the range previously determined based on real prostate tissue and manufacturer provided estimation of the inclusion stiffness ($47.7\pm17.7$~kPa for the hard inclusions and $15.9\pm5.9$~kPa for simulated tissue).

\subsection{Patient Results}

The results for the two different patient studies are presented here. The first is a summary of published results from the Prostatectomy Study. The second contains a preliminary qualitative comparison of S-WAVE images with registered whole-gland biopsy results from the Focal Therapy Study.

\subsubsection{Prostatectomy Study}

Fig.~\ref{fig:prostResults6} show examples of the comparison between the transverse absolute elasticity slice and its corresponding histopathology slice for two different patients. The excitation frequencies used to compute the average elasticity were 75~Hz, 77~Hz and 80~Hz. These images show a good correlation between the stiffness of a region and positively identified cancerous tissue from pathology. When using absolute elasticity images along with the phasor images to create a multi-parametric classifier to identify prostate cancer in the peripheral zone, an area under the receiver operating characteristic curve of $0.82\pm0.01$ was achieved~\cite{mohareri2014multi}.

\begin{figure*}[!t]
\centering
      \includegraphics[width=\textwidth, trim={0 9.5cm 7.5cm 0}, clip] {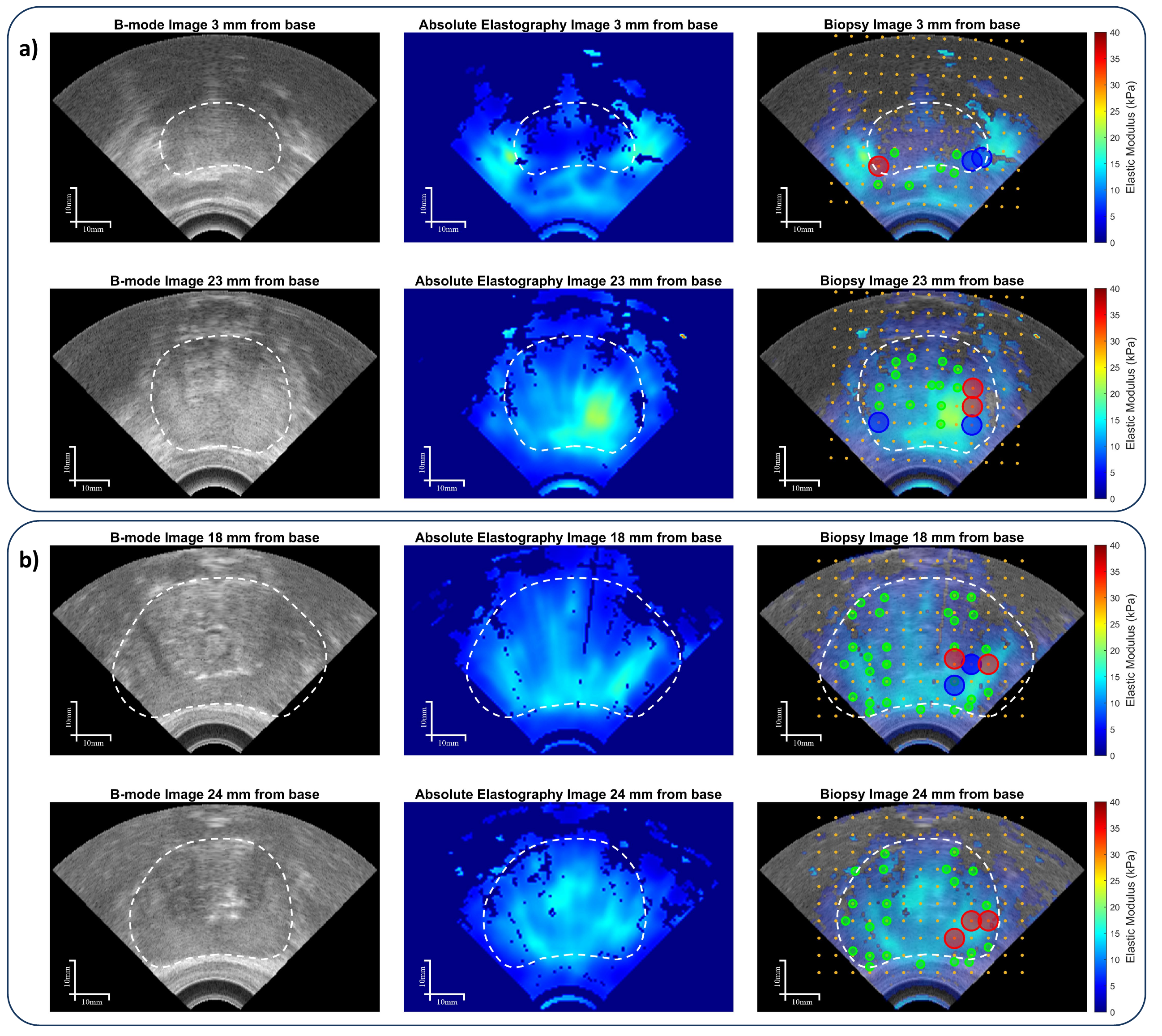}\\
 \caption{The comparison of B-mode images (left), absolute elasticity images (middle), and registered biopsy core results superimposed with the B-mode and absolute elasticity values (right) for two different slices in patient 1 (a) and patient 2 (b) from the Focal Therapy Study. For the biopsy results (right), green, blue, and red circles indicate samples that were determined to benign, positive (malignant tissue present in the core), and malignant (malignant tissue present in current plane), respectively. Orange circles are grid locations from the needle template.}
 \label{fig:FTCPpat5}
\end{figure*}

\subsubsection{Focal Therapy Study}

Biopsy cross-sections are registered with the interpolated transverse slices of the B-mode and elastography volumes using the automatic biopsy core registration technique. Fig.~\ref{fig:FTCPpat5} show the B-mode and the corresponding absolute elasticity image of two transverse slices from two patients in which cancerous regions were identified based on the pathology of the biopsy cores. The last column in these figures show the registered biopsy cores where: green, blue, and red circles indicate samples that are benign, positive (contains malignant tissue but in a different plane), and malignant (contains malignant tissue in the current plane), respectively. Orange circles identify grid locations of the needle template used to guide needles during biopsy. Due to the possibility of needle bending, cores are not always extracted from the template grid locations as can be seen for some of the cores in Fig.~\ref{fig:FTCPpat5}.

For these cases, it can be observed that cancerous identified cores are generally stiffer than surrounding tissue as measured by our elastography system. False positive results are also present such as the stiffer region seen at the bottom left of the prostate image in the second row of Fig.~\ref{fig:FTCPpat5}b, which has some corresponding negative cores. These and any other negative correlations could be due to a number of different biological phenomena, such as calcification and edema, which will be investigated in a further study. Furthermore, as the patients recruited in this study had early stage cancer, the cancerous regions are not expected to be significantly stiffer than surrounding tissue.

\section{Discussion \& Conclusion}
\label{conclusions}
This article has outlined a new ultrasound elastography platform that can be used to measure and display the 3D absolute value of the stiffness of tissue in quasi-real time. A specific implementation of the system that can be used for prostate imaging to help with identification of cancerous tissue has been described. The elasticity is computed by averaging the values obtained from several frequencies which can be applied simultaneously.  The system has been validated, using two CIRS elastography phantoms and two clinical patient studies to perform elastic modulus measurements of the prostate. The results from the phantom validation show strong agreement with MRE measurements. For the two clinical studies, a numerical and visual correlation between stiff elasticity values (as measured by our system) and cancerous regions (identified in pathology) has been also demonstrated. We are currently using this system in a new clinical study (OPTiMAL Therapy) at BC Cancer (Vancouver, Canada) investigating the use of dual strength doses in LDR-PB to minimize radiation to healthy tissues. Future work includes an extended analysis of the Focal and OPTiMAL therapy study data to further quantify the correlation between absolute elasticity as measured by this system and cancerous tissue. Furthermore, it should be noted that a simple averaging of the elasticity values from the different frequencies does not take into account variables such as 1) the quality of the phasor fit, 2) viscous components that could create a frequency dependence in the response of the measured tissue, 3) shear wave pattern artifacts at medium boundaries and 4) ``nodes'' or low amplitude displacement points that can make phasor fitting unreliable. The use of a method that takes into account all of these effects may help make the results more robust and repeatable.

% \section*{Acknowledgment}
% The preferred spelling of the word ``acknowledgment'' in American English is 
% without an ``e'' after the ``g.'' Use the singular heading even if you have 
% many acknowledgments. Avoid expressions such as ``One of us (S.B.A.) would 
% like to thank $\ldots$ .'' Instead, write ``F. A. Author thanks $\ldots$ .'' In most 
% cases, sponsor and financial support acknowledgments are placed in the 
% unnumbered footnote on the first page, not here.

\bibliographystyle{IEEEtran}
\bibliography{IEEEabrv,bibliography}   % name your BibTeX data base

\begin{thebibliography}{10}
\providecommand{\url}[1]{#1}
\csname url@samestyle\endcsname
\providecommand{\newblock}{\relax}
\providecommand{\bibinfo}[2]{#2}
\providecommand{\BIBentrySTDinterwordspacing}{\spaceskip=0pt\relax}
\providecommand{\BIBentryALTinterwordstretchfactor}{4}
\providecommand{\BIBentryALTinterwordspacing}{\spaceskip=\fontdimen2\font plus
\BIBentryALTinterwordstretchfactor\fontdimen3\font minus
  \fontdimen4\font\relax}
\providecommand{\BIBforeignlanguage}[2]{{%
\expandafter\ifx\csname l@#1\endcsname\relax
\typeout{** WARNING: IEEEtran.bst: No hyphenation pattern has been}%
\typeout{** loaded for the language `#1'. Using the pattern for}%
\typeout{** the default language instead.}%
\else
\language=\csname l@#1\endcsname
\fi
#2}}
\providecommand{\BIBdecl}{\relax}
\BIBdecl

\bibitem{lerner1988sono}
R.~M. Lerner, K.~J. Parker, J.~Holen, R.~Gramiak, and R.~C. Waag,
  ``Sono-elasticity: medical elasticity images derived from ultrasound signals
  in mechanically vibrated targets,'' in \emph{Acoustical imaging}.\hskip 1em
  plus 0.5em minus 0.4em\relax Springer, 1988, pp. 317--327.

\bibitem{witters2009nliverfib}
P.~Witters, K.~De~Boeck, L.~Dupont, M.~Proesmans, F.~Vermeulen, R.~Servaes,
  C.~Verslype, W.~Laleman, F.~Nevens, I.~Hoffman \emph{et~al.}, ``Non-invasive
  liver elastography (fibroscan) for detection of cystic fibrosis-associated
  liver disease,'' \emph{Journal of Cystic Fibrosis}, vol.~8, no.~6, pp.
  392--399, 2009.

\bibitem{castera2005liver}
L.~Cast{\'e}ra, J.~Vergniol, J.~Foucher, B.~Le~Bail, E.~Chanteloup, M.~Haaser,
  M.~Darriet, P.~Couzigou, and V.~de~L{\'e}dinghen, ``Prospective comparison of
  transient elastography, fibrotest, apri, and liver biopsy for the assessment
  of fibrosis in chronic hepatitis c,'' \emph{Gastroenterology}, vol. 128,
  no.~2, pp. 343--350, 2005.

\bibitem{warner2011kidney}
L.~Warner, M.~Yin, K.~J. Glaser, J.~A. Woollard, C.~A. Carrascal, M.~J. Korsmo,
  J.~A. Crane, R.~L. Ehman, and L.~O. Lerman, ``Noninvasive in vivo assessment
  of renal tissue elasticity during graded renal ischemia using mr
  elastography,'' \emph{Investigative radiology}, vol.~46, no.~8, p. 509, 2011.

\bibitem{edwards2020use}
C.~Edwards, E.~Cavanagh, S.~Kumar, V.~Clifton, and D.~Fontanarosa, ``The use of
  elastography in placental research--a literature review,'' \emph{Placenta},
  vol.~99, pp. 78--88, 2020.

\bibitem{correas2015prostate}
J.-M. Correas, A.-M. Tissier, A.~Khairoune, V.~Vassiliu, A.~M{\'e}jean,
  O.~H{\'e}l{\'e}non, R.~Memo, and R.~G. Barr, ``Prostate cancer: diagnostic
  performance of real-time shear-wave elastography,'' \emph{Radiology}, vol.
  275, no.~1, pp. 280--289, 2015.

\bibitem{arroyo2018breast}
J.~Arroyo, A.~C. Saavedra, J.~Guerrero, P.~Montenegro, J.~Aguilar, J.~A. Pinto,
  J.~Lobo, T.~Salcudean, R.~Lavarello, and B.~Casta{\~n}eda, ``Breast
  elastography: Identification of benign and malignant cancer based on absolute
  elastic modulus measurement using vibro-elastography,'' in \emph{Medical
  Imaging 2018: Image Perception, Observer Performance, and Technology
  Assessment}, vol. 10577.\hskip 1em plus 0.5em minus 0.4em\relax International
  Society for Optics and Photonics, 2018, p. 105771E.

\bibitem{sang2017accuracy}
L.~Sang, X.-m. Wang, D.-y. Xu, and Y.-f. Cai, ``Accuracy of shear wave
  elastography for the diagnosis of prostate cancer: A meta-analysis,''
  \emph{Scientific reports}, vol.~7, no.~1, pp. 1--8, 2017.

\bibitem{eskandari2008viscoelastic}
H.~Eskandari, S.~E. Salcudean, and R.~Rohling, ``Viscoelastic parameter
  estimation based on spectral analysis,'' \emph{Ultrasonics, Ferroelectrics
  and Frequency Control, IEEE Transactions on}, vol.~55, no.~7, pp. 1611--1625,
  2008.

\bibitem{sigrist2017ultrasound}
R.~M. Sigrist, J.~Liau, A.~El~Kaffas, M.~C. Chammas, and J.~K. Willmann,
  ``Ultrasound elastography: review of techniques and clinical applications,''
  \emph{Theranostics}, vol.~7, no.~5, p. 1303, 2017.

\bibitem{ophir1991elastography}
J.~Ophir, I.~Cespedes, H.~Ponnekanti, Y.~Yazdi, and X.~Li, ``Elastography: a
  quantitative method for imaging the elasticity of biological tissues,''
  \emph{Ultrasonic imaging}, vol.~13, no.~2, pp. 111--134, 1991.

\bibitem{lorenz1999new}
A.~Lorenz, H.-J. Sommerfeld, M.~Garcia-Schurmann, S.~Philippou, T.~Senge, and
  H.~Ermert, ``A new system for the acquisition of ultrasonic multicompression
  strain images of the human prostate in vivo,'' \emph{Ultrasonics,
  Ferroelectrics and Frequency Control, IEEE Transactions on}, vol.~46, no.~5,
  pp. 1147--1154, 1999.

\bibitem{wildeboer2020automated}
R.~R. Wildeboer, C.~K. Mannaerts, R.~J. van Sloun, L.~Bud{\"a}us, D.~Tilki,
  H.~Wijkstra, G.~Salomon, and M.~Mischi, ``Automated multiparametric
  localization of prostate cancer based on b-mode, shear-wave elastography, and
  contrast-enhanced ultrasound radiomics,'' \emph{European radiology}, vol.~30,
  no.~2, pp. 806--815, 2020.

\bibitem{dai2021correlation}
W.-B. Dai, J.~Xu, B.~Yu, L.~Chen, Y.~Chen, and J.~Zhan, ``Correlation of
  stiffness of prostate cancer measured by shear wave elastography with grade
  group: A preliminary study,'' \emph{Ultrasound in Medicine \& Biology},
  vol.~47, no.~2, pp. 288--295, 2021.

\bibitem{ji2019stiffness}
Y.~Ji, L.~Ruan, W.~Ren, G.~Dun, J.~Liu, Y.~Zhang, and Q.~Wan, ``Stiffness of
  prostate gland measured by transrectal real-time shear wave elastography for
  detection of prostate cancer: a feasibility study,'' \emph{The British
  journal of radiology}, vol.~92, no. 1097, p. 20180970, 2019.

\bibitem{zhang2008quantitative}
M.~Zhang, P.~Nigwekar, B.~Castaneda, K.~Hoyt, J.~V. Joseph, A.~di~Sant'Agnese,
  E.~M. Messing, J.~G. Strang, D.~J. Rubens, and K.~J. Parker, ``Quantitative
  characterization of viscoelastic properties of human prostate correlated with
  histology,'' \emph{Ultrasound in medicine \& biology}, vol.~34, no.~7, pp.
  1033--1042, 2008.

\bibitem{zhai2010acoustic}
L.~Zhai, J.~Madden, W.-C. Foo, M.~L. Palmeri, V.~Mouraviev, T.~J. Polascik, and
  K.~R. Nightingale, ``Acoustic radiation force impulse imaging of human
  prostates ex vivo,'' \emph{Ultrasound in medicine \& biology}, vol.~36,
  no.~4, pp. 576--588, 2010.

\bibitem{ahmad2013transrectal}
S.~Ahmad, R.~Cao, T.~Varghese, L.~Bidaut, and G.~Nabi, ``Transrectal
  quantitative shear wave elastography in the detection and characterisation of
  prostate cancer,'' \emph{Surgical endoscopy}, vol.~27, no.~9, pp. 3280--3287,
  2013.

\bibitem{gennisson2015}
J.-L. Gennisson, J.~Provost, T.~Deffieux, C.~Papadacci, M.~Imbault, M.~Pernot,
  and M.~Tanter, ``4-d ultrafast shear-wave imaging,'' \emph{Ultrasonics,
  Ferroelectrics, and Frequency Control, IEEE Transactions on}, vol.~62, no.~6,
  pp. 1059--1065, 2015.

\bibitem{nightingale2011acoustic}
K.~Nightingale, ``Acoustic radiation force impulse (arfi) imaging: a review,''
  \emph{Current medical imaging}, vol.~7, no.~4, pp. 328--339, 2011.

\bibitem{baghani2012real}
A.~Baghani, H.~Eskandari, W.~Wang, D.~Da~Costa, M.~N. Lathiff, R.~Sahebjavaher,
  S.~Salcudean, and R.~Rohling, ``Real-time quantitative elasticity imaging of
  deep tissue using free-hand conventional ultrasound,'' in \emph{Medical Image
  Computing and Computer-Assisted Intervention--MICCAI 2012}.\hskip 1em plus
  0.5em minus 0.4em\relax Springer, 2012, pp. 617--624.

\bibitem{schneider2012remote}
C.~Schneider, A.~Baghani, R.~Rohling, and S.~Salcudean, ``Remote ultrasound
  palpation for robotic interventions using absolute elastography,'' in
  \emph{International Conference on Medical Image Computing and
  Computer-Assisted Intervention}.\hskip 1em plus 0.5em minus 0.4em\relax
  Springer, 2012, pp. 42--49.

\bibitem{honarvar2014vibro}
M.~Honarvar, S.~E. Salcudean, and R.~N. Rohling, ``Vibro-elastography: direct
  fem inversion of the shear wave equation without the local homogeneity
  assumption,'' in \emph{Medical Imaging 2014: Ultrasonic Imaging and
  Tomography}, vol. 9040.\hskip 1em plus 0.5em minus 0.4em\relax International
  Society for Optics and Photonics, 2014, p. 904003.

\bibitem{manduca1996local}
A.~Manduca, R.~Muthupillai, P.~Rossman, J.~F. Greenleaf, and R.~L. Ehman,
  ``Local wavelength estimation for magnetic resonance elastography,'' in
  \emph{Image Processing, 1996. Proceedings., International Conference on},
  vol.~3.\hskip 1em plus 0.5em minus 0.4em\relax IEEE, 1996, pp. 527--530.

\bibitem{asbach2008assessment}
P.~Asbach, D.~Klatt, U.~Hamhaber, J.~Braun, R.~Somasundaram, B.~Hamm, and
  I.~Sack, ``Assessment of liver viscoelasticity using multifrequency mr
  elastography,'' \emph{Magnetic Resonance in Medicine}, vol.~60, no.~2, pp.
  373--379, 2008.

\bibitem{deffieux2009shear}
T.~Deffieux, G.~Montaldo, M.~Tanter, and M.~Fink, ``Shear wave spectroscopy for
  in vivo quantification of human soft tissues visco-elasticity,''
  \emph{Medical Imaging, IEEE Transactions on}, vol.~28, no.~3, pp. 313--322,
  2009.

\bibitem{zeng2020three}
Q.~Zeng, M.~Honarvar, C.~Schneider, S.~K. Mohammad, J.~Lobo, E.~H. Pang, K.~T.
  Lau, C.~Hu, J.~Jago, S.~R. Erb \emph{et~al.}, ``Three-dimensional
  multi-frequency shear wave absolute vibro-elastography (3d s-wave) with a
  matrix array transducer: Implementation and preliminary in vivo study of the
  liver,'' \emph{IEEE Transactions on Medical Imaging}, vol.~40, no.~2, pp.
  648--660, 2020.

\bibitem{abeysekera2017swave}
J.~M. Abeysekera, M.~Ma, M.~Pesteie, J.~Terry, D.~Pugash, J.~A. Hutcheon,
  C.~Mayer, L.~Lampe, S.~Salcudean, and R.~Rohling, ``Swave imaging of
  placental elasticity and viscosity: proof of concept,'' \emph{Ultrasound in
  Medicine \& Biology}, vol.~43, no.~6, pp. 1112--1124, 2017.

\bibitem{shao2021breast}
Y.~Shao, H.~Hashemi, P.~Gordon, L.~Warren, Z.~J. Wang, R.~Rohling, and
  T.~Salcudean, ``Breast cancer detection using multimodal time series features
  from ultrasound shear wave absolute vibro-elastography,'' \emph{IEEE Journal
  of Biomedical and Health Informatics}, 2021.

\bibitem{lobo2015prostate}
J.~Lobo, A.~Baghani, H.~Eskandari, S.~Mahdavi, R.~Rohling, L.~Goldernberg,
  W.~J. Morris, and S.~Salcudean, ``Prostate vibro-elastography:
  Multi-frequency 1d over 3d steady-state shear wave imaging for quantitative
  elastic modulus measurement,'' in \emph{Ultrasonics Symposium (IUS), 2015
  IEEE International}.\hskip 1em plus 0.5em minus 0.4em\relax IEEE, 2015, pp.
  1--4.

\bibitem{eskandari2011bandpass}
H.~Eskandari, O.~Goksel, S.~Salcudean, and R.~Rohling, ``Bandpass sampling of
  high-frequency tissue motion,'' \emph{Ultrasonics, Ferroelectrics and
  Frequency Control, IEEE Transactions on}, vol.~58, no.~7, pp. 1332--1343,
  2011.

\bibitem{baghani2010high}
A.~Baghani, A.~Brant, S.~Salcudean, and R.~Rohling, ``A high-frame-rate
  ultrasound system for the study of tissue motions,'' \emph{Ultrasonics,
  Ferroelectrics and Frequency Control, IEEE Transactions on}, vol.~57, no.~7,
  pp. 1535--1547, 2010.

\bibitem{sahebjavaher2013transperineal}
R.~S. Sahebjavaher, A.~Baghani, M.~Honarvar, R.~Sinkus, and S.~E. Salcudean,
  ``Transperineal prostate mr elastography: initial in vivo results,''
  \emph{Magnetic Resonance in Medicine}, vol.~69, no.~2, pp. 411--420, 2013.

\bibitem{mohareri2014multi}
O.~Mohareri, A.~Ruszkowski, J.~Lobo, J.~Ischia, A.~Baghani, G.~Nir,
  H.~Eskandari, E.~Jones, L.~Fazli, L.~Goldenberg \emph{et~al.},
  ``Multi-parametric 3d quantitative ultrasound vibro-elastography imaging for
  detecting palpable prostate tumors,'' in \emph{Medical Image Computing and
  Computer-Assisted Intervention--MICCAI 2014}.\hskip 1em plus 0.5em minus
  0.4em\relax Springer, 2014, pp. 561--568.

\bibitem{zahiri2006motion}
R.~Zahiri-Azar and S.~E. Salcudean, ``Motion estimation in ultrasound images
  using time domain cross correlation with prior estimates,'' \emph{Biomedical
  Engineering, IEEE Transactions on}, vol.~53, no.~10, pp. 1990--2000, 2006.

\bibitem{kolsky1964stress}
H.~Kolsky, ``Stress waves in solids,'' \emph{Journal of sound and Vibration},
  vol.~1, no.~1, pp. 88--110, 1964.

\bibitem{knutsson1994local}
H.~Knutsson, C.~Westin, and G.~Granlund, ``Local multiscale frequency and
  bandwidth estimation,'' in \emph{Image Processing, 1994. Proceedings.
  ICIP-94., IEEE International Conference}, vol.~1.\hskip 1em plus 0.5em minus
  0.4em\relax IEEE, 1994, pp. 36--40.

\bibitem{baghani2011travelling}
A.~Baghani, S.~Salcudean, M.~Honarvar, R.~Sahebjavaher, R.~Rohling, and
  R.~Sinkus, ``Travelling wave expansion: A model fitting approach to the
  inverse problem of elasticity reconstruction,'' \emph{Medical Imaging, IEEE
  Transactions on}, vol.~30, no.~8, pp. 1555--65, Aug. 2011.

\bibitem{nir2013registration}
G.~Nir and S.~E. Salcudean, ``Registration of whole-mount histology and
  tomography of the prostate using particle filtering,'' in \emph{Proc. SPIE},
  vol. 8676, 2013, pp. 86\,760E1--9.

\bibitem{aleef2022}
T.~A. Aleef, Q.~Zeng, W.~J. Morris, S.~S. Mahdavi, and S.~E. Salcudean,
  ``Registration of trans-perineal templatemapping biopsy cores to
  volumetricultrasound,'' \emph{International Journal of Computer Assisted
  Radiology and Surgery}, 2022.

\end{thebibliography}

\end{document}